\documentclass[manuscript]{aastex}

\slugcomment{to appear in ApJ}

\shorttitle{Extended Main-Sequence Turn-off Clusters}
\shortauthors{Keller, Mackey, Da Costa}

\begin{document}


\title{The Extended Main-Sequence Turn-off Clusters of the Large Magellanic Cloud --- Missing links in Globular Cluster Evolution}


\author{Stefan C.\ Keller, A.\ Dougal Mackey and Gary S.\ Da Costa}
\affil{Research School of Astronomy and Astrophysics, Australian National University, \\Mt. Stromlo Observatory, Cotter Rd. Weston ACT 2611 Australia.}
\email{stefan@mso.anu.edu.au}



\begin{abstract}
Recent observations of intermediate age (1 -- 3 Gyr) massive star clusters in the Large Magellanic Cloud (LMC) have revealed that the majority possess bifurcated or extended main-sequence turn-off (EMSTO) morphologies. This effect can be understood to arise from subsequent star formation amongst the stellar population with age differences between constituent stars amounting to 50 -- 300 Myr. Age spreads of this order are similarly invoked to explain the light element abundance variations witnessed in ancient globular clusters. In this paper we explore the proposition that the clusters exhibiting the EMSTO phenomenon are a general phase in the evolution of massive clusters, one that naturally leads to the particular chemical properties of the ancient globular cluster population. 

We show that the isolation of EMSTO clusters to intermediate ages is the consequence of observational selection effects. In our proposed scenario, the EMSTO phenomenon is identical to that which establishes the light element abundance variations that are ubiquitous in the ancient globular cluster population. Our scenario makes a strong prediction: EMSTO clusters will exhibit abundance variations in the light elements characteristic of the ancient GC population. 
\end{abstract}



\section{Introduction}

A number of intermediate age (1-3 Gyr) clusters in the Large Magellanic Cloud (LMC) exhibit a striking peculiarity, first noticed by \citet{Bertelli03}, namely a main-sequence turnoff (MSTO) that is bifurcated or excessively broad in luminosity relative to photometric uncertainties. In the case of NGC 1846 \citet{Mackey07} show that the cluster exhibits two distinct MSTOs. Extended MSTOs (EMSTOs) were subsequently reported in the clusters NGC 1806 and 1783 \citep{Mackey08a, Goudfrooij09} and the SMC cluster NGC 419 \citet{Glatt08b}. \citet{Milone09} examined the MSTO morphology of 16 intermediate age LMC clusters and found that 11 ($\sim70$\%) exhibit EMSTOs. In contrast to the MSTO, the red giant branch (RGB) and main-sequence (MS) are narrow and well defined. This indicates that the cluster stellar population(s) possesses a low range in metallicity. The aforementioned studies also demonstrate that the morphology of the color-magnitude diagram (CMD) at the MSTO can not be explained by field star contamination or cluster binary stars. Instead, the studies conclude, the clusters are composed of stars with ages spanning a range of up to 300Myr. This at odds with the canonical picture of a star cluster as a coeval single stellar population. 

\citet{Bastian09} provided an alternative explanation for the presence of EMSTOs based on the effects of stellar rotation on the observed CMD. This study shows that moderate rotation ($\omega=0.4\omega_{c}$, where $\omega_c$ is the critical break-up velocity) amongst M$>1.2$M$_{\odot}$ stars can mimic the observed CMD of EMSTO clusters. The effects of rapid rotation are evident in the young ($\sim 10^{7}$yr) cluster system of the MCs in the large fraction of Be stars (stars rotating at considerable fractions of $\omega_c$; \citeauthor{Keller99} \citeyear{Keller99}). However, such rapid rotation is not seen amongst lower mass A-type stars \citep{Wolff82, Keller04} that populate the MSTO of older ($\sim$ 10$^8$yr old) clusters such as NGC 1866 \citep{Testa99}. \citet{Girardi09} consider a rotational scenario in the context of the SMC EMSTO cluster NGC 419. This cluster exhibits twin red clumps. Such dual red clumps require a dispersion in the mass of the H-exhausted core of $\sim$0.2M$_{\odot}$ (or $\sim 15$\%). It is unlikely that rotation can produce this magnitude of effect \citep{Girardi09}. In addition, it is concluded by \citet{Girardi10} that the effect of convective core overshoot is similarly unable to account of the MSTO, rather they conclude that an age spread is the most likely cause.

Assuming that age is responsible for the MSTO morphology, the observations suggest an age range rather than a strong age bimodality for most clusters \citep{Goudfrooij09}. Age ranges of between 50-300Myr are observed in the LMC sample \citep{Milone09}. However, it must be borne in mind that existing photometry, largely of short exposure or `snapshot' quality, will limit our ability to discern small age spreads and clusters in which the distribution of ages is highly skewed towards one extreme of the age range. It is also not clear if the MSTOs of clusters that apparently exhibit a broad age range will show age bimodality under closer observational scrutiny.

By dividing the MSTO into brighter and fainter branches, \citeauthor{Milone09} shows that NGC 1846, 1806 and 1751 possess a dominant brighter (younger) population that accounts for $\sim70$\% of the cluster by number. Furthermore, the \citet{Goudfrooij09} study of NGC 1846 shows that the younger population is significantly more centrally concentrated than the older population, although this result was not recovered by \citet{Mackey07}. In general, however, the MSTO morphology for the known EMSTO clusters is not clearly bimodal.

Other than rotation, three scenarios are discussed in the literature to explain the formation of EMSTO clusters. First is the scenario that EMSTO clusters result from the merger of binary star clusters \citep{Mackey07}. The MCs are known to contain binary clusters \citep{Bhatia88, Dieball00}. However, the clusters that constitute binary systems are essentially coeval with age differences of $\sim$few$\times10^{7}$yr. \citet{Efremov98} find that the average age difference between cluster pairs in the LMC increases with their angular separation. Hence a larger age difference could arise from clusters that formed at large spatial separation in a giant molecular cloud (GMC). However, as pointed out by \citet{Goudfrooij09} the spatial separation required to explain a $\sim$200Myr age difference is many times larger than the scale of a typical GMC. Similarly, it is unlikely that two clusters that formed in isolation and then subsequently drawn together should possess such similar metallicity. Hence formation via the agglomeration of two clusters is unlikely \citep{Mackey08a}.

\citet{Bekki09} explore a second formation scenario in which an initial star cluster interacts and merges with a GMC, undergoes an episode of star formation, to form a composite star cluster with two distinct constituent populations. The advantages of this scenario are that: it presents a large reservoir of gas from which a large number of second generation stars can be formed, and the second generation is naturally more centrally condensed, in line with observations of \citet{Goudfrooij09}.

The final scenario for the formation of EMSTO clusters is that the age spread arises from insitu star formation\citep{Goudfrooij09}. The challenges for this scenario are: how does a small potential, such as those of EMSTO clusters, manage to retain gas for secondary or protracted star formation in the face of mechanical feedback that results from the course of stellar evolution (i.e. SNII)? How can a secondary population of up to $\sim$70\% arise in such a scenario? 

The insitu formation scenario for EMSTO clusters parallels that of the formation of multiple populations in Galactic globular clusters (GGCs). The origin of the observed star-to-star light element abundance inhomogeneities in the GGCs has been the topic of an extensive literature (see e.g.\ \citeauthor{Carretta10} \citeyear{Carretta10} for a review). The observed presence of star-to-star variations from the RGB to the MS argues that the abundance variations are not due to mixing and/or chemical processes internal to each star. Rather, it strongly argues that the abundance inhomogeneities are due to the incorporation of chemically processed material derived from previous generation(s) (the self-pollution scenario, first proposed by \citeauthor{Cottrell81} \citeyear{Cottrell81}). The presence of multiple populations in GGCs appears ubiquitous \citep{Carretta10} and is also apparent in extragalactic globular cluster systems (Fornax; \citeauthor{Letarte06} \citeyear{Letarte06}, and the ancient GCs of the LMC; \citeauthor{Mucciarelli09} \citeyear{Mucciarelli09}). 

There are a number of intriguing similarities between the properties of the multiple populations of ancient GCs and those evident in the EMSTO clusters: firstly the age range of $\sim$100-300 Myr for formation of the first and second generations in EMSTO clusters is of the same order as that required by of some of the processes (e.g.\ AGB stars) postulated to generate the processed material in the ancient GCs; secondly some EMSTO clusters are dominated by a second generation, similar to that evident in ancient GCs \citep{DAntona08}.

In this paper we explore the proposition that the EMSTO phenomenon is related to the phenomenon of multiple populations evident in ancient GCs. In Section 2 we demonstrate that there are strong observational selection biases that restrict our ability to detect EMSTO clusters to a narrow age range that overlies the intermediate age clusters of the LMC. In Section 3 we show that the EMSTO clusters tend to be the most spatially extended clusters in their age range, and furthermore that the strength of the EMSTO is correlated with the degree of spatial extension. In Section 4 we argue for a scenario in which evolution shapes the luminosity function of the massive cluster population such that those that have experienced the EMSTO phenomenon remain to form the ancient GC population. In this scenario, the EMSTO clusters of the LMC are the `missing-links' of GC evolution --- it is in the EMSTO phenomenon that the peculiar GC abundance anomalies are established.

\section{Extended MSTO selection effects in intermediate age clusters}

The chemical inhomogeneity of light elements is ubiquitous in all GC systems studied to date. If the extended MSTO phenomenon is related to the production of these chemical inhomogeneities why then do we only see extended MSTOs in clusters in a particular age range? In this section we demonstrate that there are strong observational selection biases against seeing clusters with multiple age populations outside this age range.

\begin{figure}
\begin{center}
\includegraphics[width=120mm]{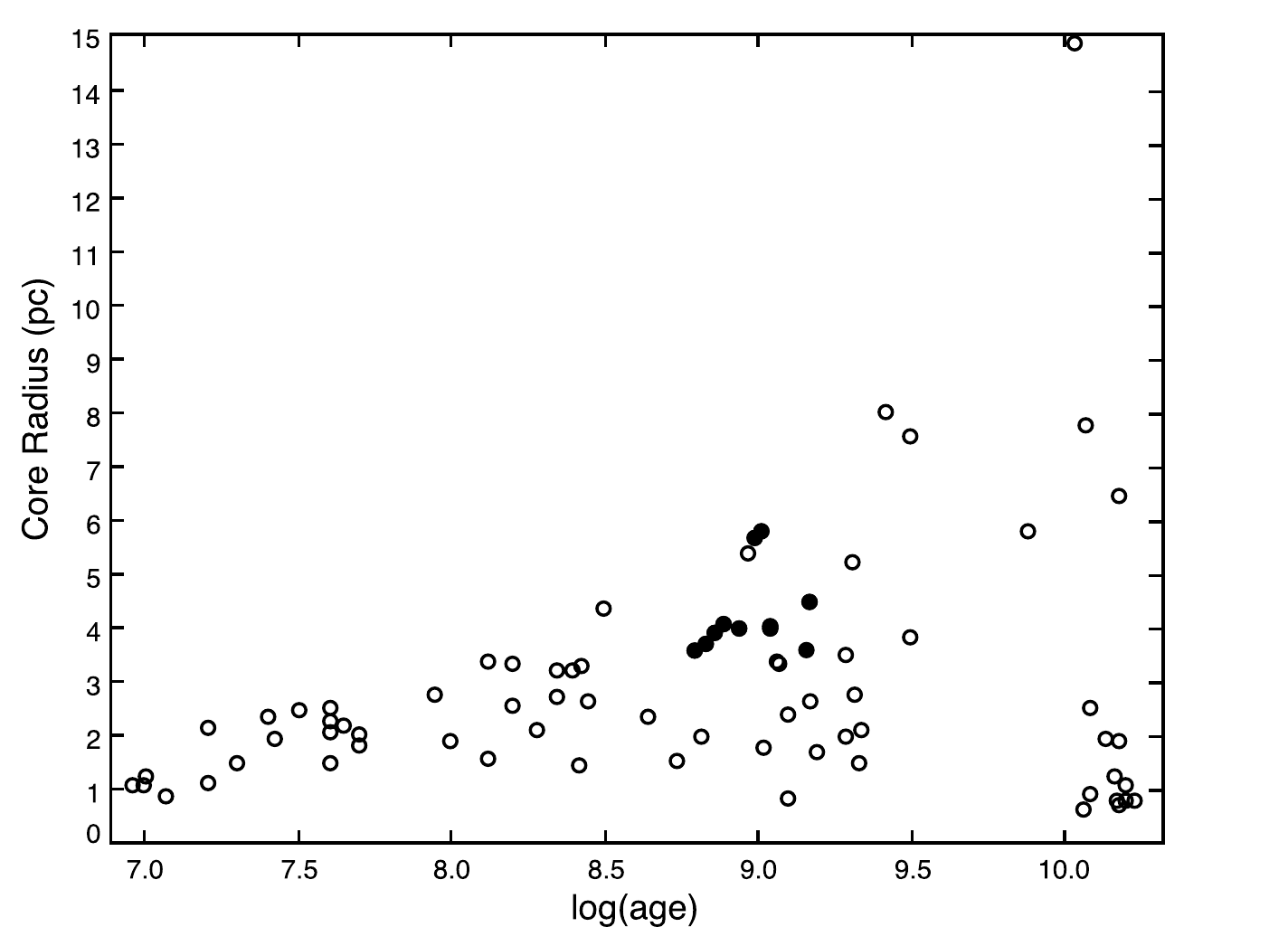}
\caption{r$_c$ versus log age for the clusters of the LMC with literature data. The {\it{known}} EMSTO clusters are indicated as solid points.}\label{figure:rc_age}
\end{center}
\end{figure}

In Figure \ref{figure:rc_age} we show the cluster core radii for LMC clusters with a variety of ages. These data are taken from the tabulation of \citet{Mackey03}. Where possible we have updated these values with refined ages and spatial parameters from recent $ACS$ imaging \citep{Mackey08a}. In filled circles we indicate those clusters that are known to be EMSTO clusters from \citet{Milone09}. \citeauthor{Milone09} finds that between 600 Myrs and 1.5 Gyrs EMSTOs account for $70\pm25$\% of clusters (11 out of 16) examined with sufficient photometric accuracy. It was noted by \citet{Bertelli03} that the 2 Gyr old cluster NGC 2173 exhibits a MSTO that can be understood as due to the presence of two populations 300 Myr apart. In the case of 0.8 Gyr old cluster NGC 1868, \citet{Santiago02} note a possible old MSTO at $\sim$3 Gyr at the level of a 5-12\% contribution to the cluster.

\begin{figure}
\begin{center}
\includegraphics[width=120mm]{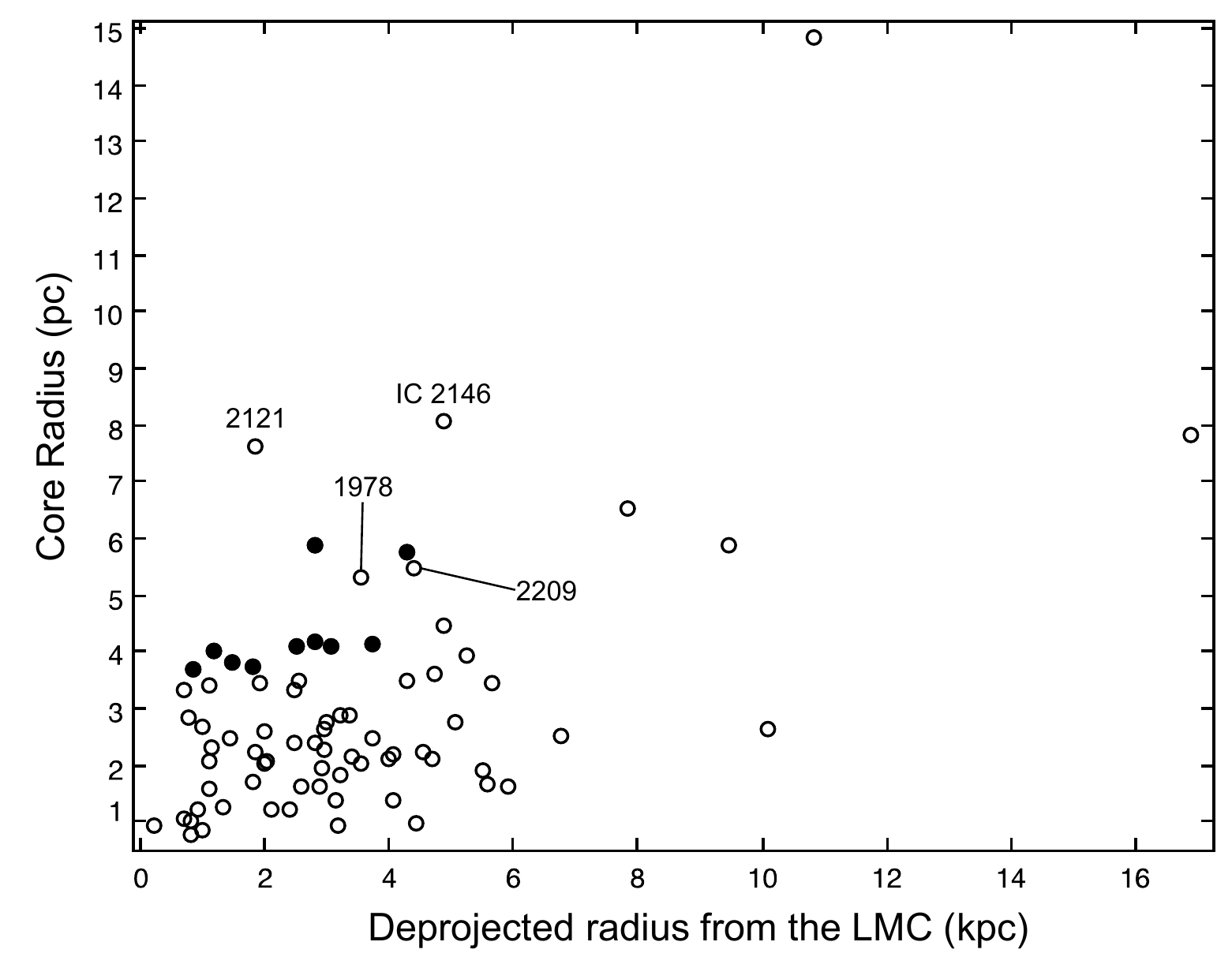}
\caption{r$_c$ versus deprojected distance from the kinematic centre of the LMC.}\label{figure:rc_rprime}
\end{center}
\end{figure}

It is particularly striking that the EMSTO clusters inhabit the upper envelope of the r$_c$ -- age distribution. This is even more evident when we control for the dependency of r$_c$ with distance from the kinematic centre of the LMC ($R$\footnote{the positions of the LMC clusters have been deprojected to the in-plane LMC distance via the kinematic LMC model as described by \citet{vanderMarel02} and updated by \citet{Olsen07}.}). It has been noted that there is a significant proximity effect present in the distribution of core radii of LMC clusters \citet{Mackey04}. This has the sense that inner clusters are generally more concentrated than those in the outskirts of the LMC (see Figure \ref{figure:rc_rprime}). 

\begin{figure}
\begin{center}
\includegraphics[width=120mm]{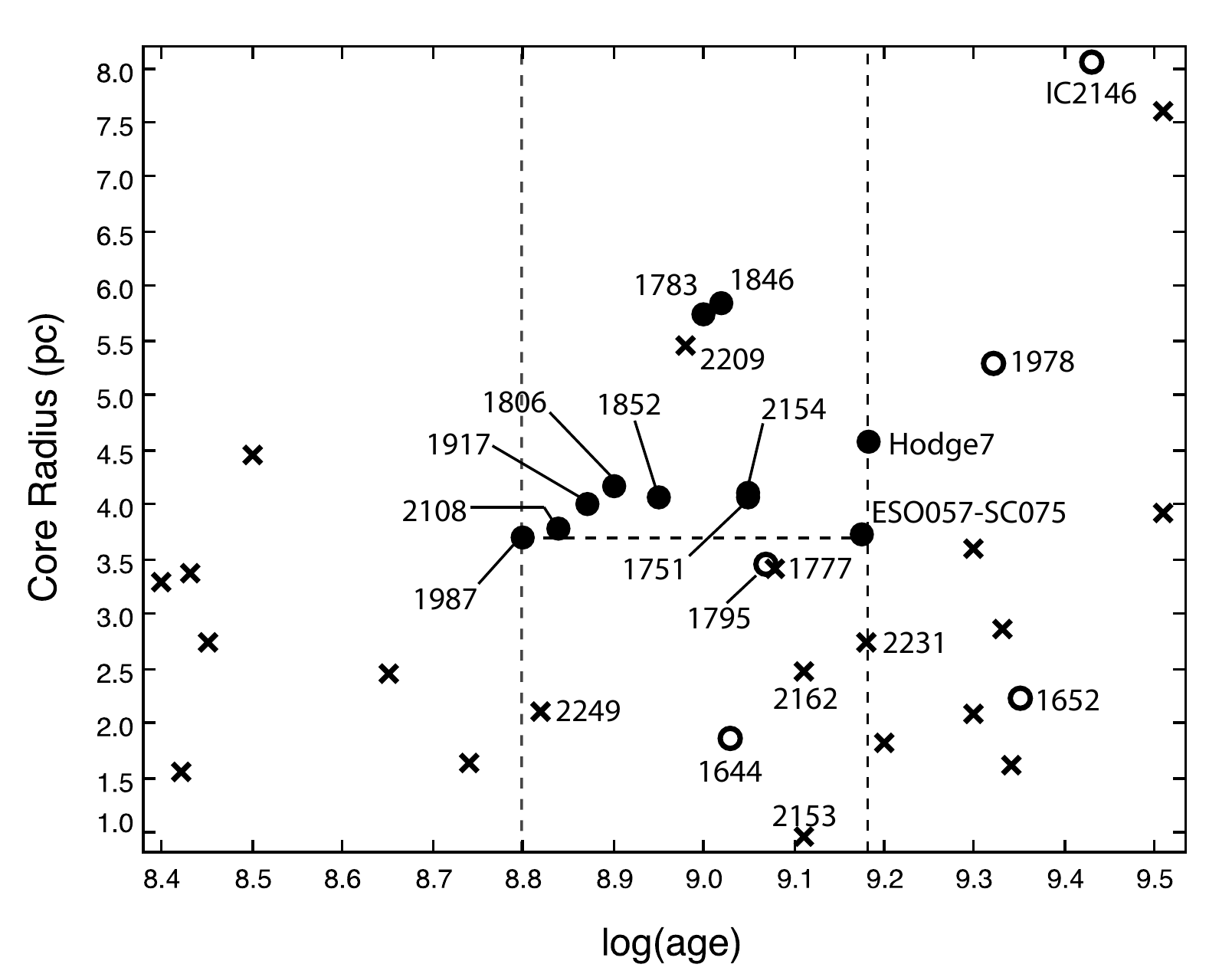}
\caption{An enlargement of the age range spanned by the currently known EMSTO clusters. Solid symbols represent the location of known EMSTO clusters from \citet{Milone09}, open circles show clusters examined by \citeauthor{Milone09} and do not possess EMSTOs. The crosses show clusters that were not included in the \citeauthor{Milone09} study. In the region of parameter space inhabited by the EMSTO clusters, 11 clusters examined exhibit the EMSTO phenomenon. Amongst clusters with smaller core radii, none of the two clusters examined with sufficient photometric accuracy exhibit the EMSTO phenomenon.}
\label{figure:zoom}
\end{center}
\end{figure}

In Figure \ref{figure:zoom} we show and expanded view of the age range covered by known EMSTO clusters in the LMC. Over the extent of age, r$_c$, and $R$ parameter space inhabited by the currently known EMSTOs there is no cluster that is known {\it{not}} to exhibit the EMSTO phenomenon. NGC 2209 lies just outside this volume making it a strong candidate for possessing the EMSTO phenomenon. No detailed photometry exists for this cluster, however.  We are not aware of any selection effects that would otherwise preclude us from finding EMSTO clusters of small core radii. Even in the most compact of cluster studied to date in this age range, NGC 1644, the HST $ACS$ photometry utilised in the \citet{Milone09} study is not sufficiently perturbed by crowding. Unfortunately however, the number of clusters that have been so examined with core radii less than the smallest EMSTO cluster (NGC 1987 with r$_c = 3.68$ pc; see Fig.\ \ref{figure:zoom}) is only two. Neither is seen to possess the EMSTO phenomenon. This should be contrasted with the 11 clusters larger than this core radius in which 11 are found to possess EMSTOs.

We now investigate the selection effects present as a function of cluster age. We will consider that the EMSTO phenomenon is based on the presence of an age spread in the star formation history of the cluster. At ages younger than the age spread exhibited in the intermediate age EMSTO clusters, we can not observe the phenomenon by definition --- the second generation has not yet formed. As we move to older ages, the relative fraction of the cluster lifetime that the age spread represents diminishes. Hence the offset in magnitude between the younger and older age MSTOs on the CMD will become smaller and the EMSTO effect will become increasingly harder to distinguish photometrically.

In order to quantify the selection effects acting against resolving the EMSTO phenomenon in older (i.e. log age $>$ 9.2) clusters, we have conducted a series of simulations as follows. We firstly assume that the second generation (SG) population is a distinct, and instantaneous, secondary star formation event and that this occurs 200 Myr after the formation of the first generation (FG). 200 Myr is chosen as it is the median of the age spreads inferred to exist in the LMC EMSTO cluster sample studied to date \citep[][their  Table 3]{Milone09}. The FG and SG populations are represented in equal proportions: this is an ``optimistic'' scenario for EMSTO visibility since typically the FG and SG populations are less bimodal but rather suggest an age spread. Where the SG and FG populations are clear the ratio of SG to FG appears to be around 70\% \citep{Milone09}.

\begin{figure}
\begin{center}
\includegraphics[width=104mm]{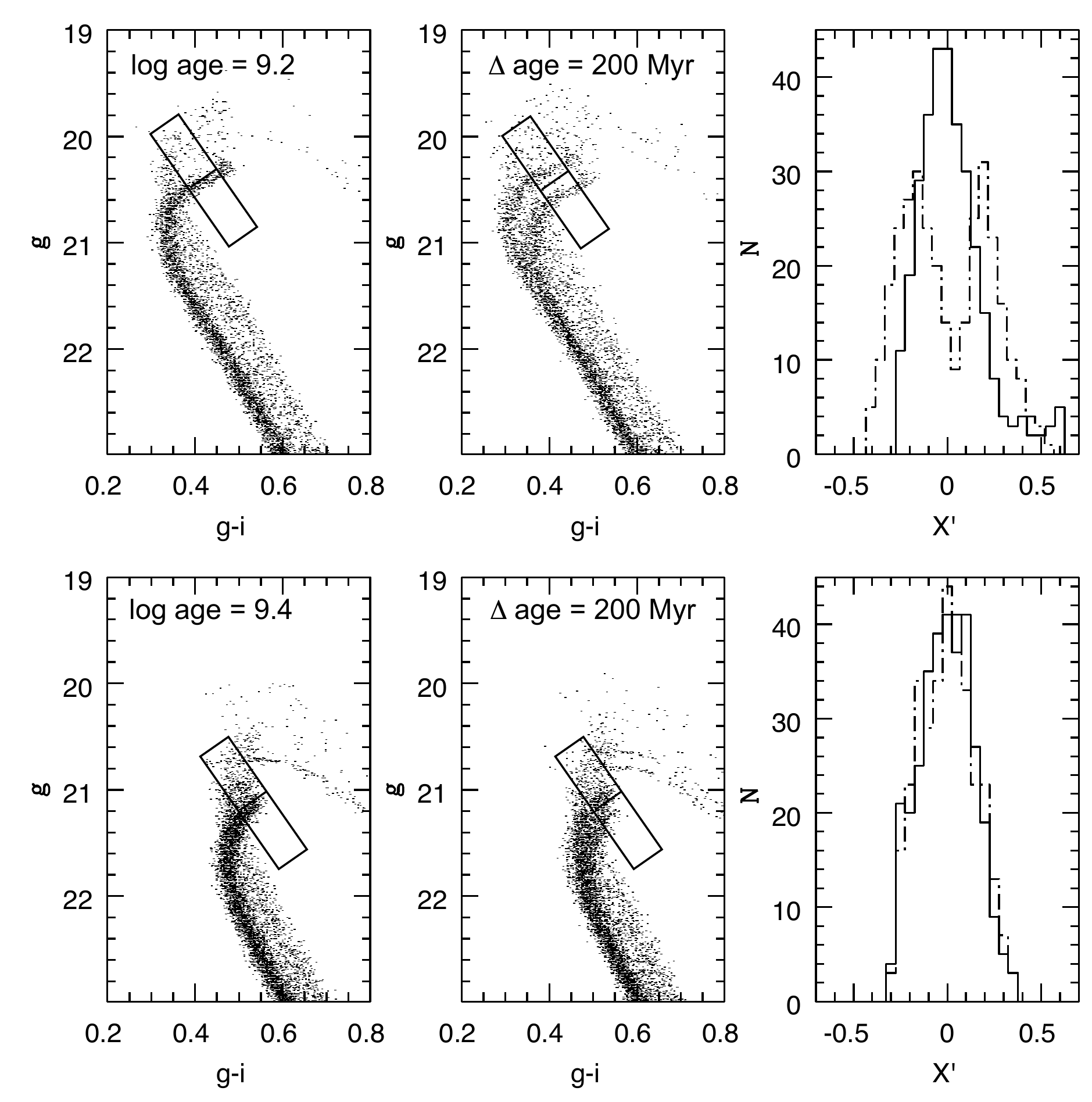}
\caption{{\bf{Top}}: Simulated color-magnitude diagrams for log age = 9.2 in the case of a single age populations (left), and for an age difference of 200 Myr (center). The rectangular box shows the MSTO region considered for our visibility analysis. The central line in this rectangle shows $X^{\prime} = 0$, positive $X^{\prime}$ are towards brighter magnitudes. The right-most panel shows the histogram of the population at the main-sequence turn-off for the single age population (solid line) and the dual population (dot-dashed line, with first and second generation stars in equal numbers). The case of the log age = 9.2, $\Delta$ age = 200 Myr is easily distinguished from the single age population. In the calculation of our visibility statistics we compute multiple realisations of these populations, see text for details. {\bf{Bottom}}: As above but for log age = 9.4. Here an age difference of 200 Myr is less distinct from the single age population.}
\label{figure:CMD}
\end{center}
\end{figure}

We then populate isochrones of appropriate ages \citep[from][]{Girardi02} to achieve a cluster of total absolute magnitude M$_V$=-7.14 (the mean M$_V$ of the extant EMSTO sample of the LMC). Photometric uncertainties appropriate for typical `snapshot' $ACS$ photometry \citep[as described in][]{Milone09} were then imposed on the model CMD (this uncertainty in the photometry is of order $\pm0.01$ magnitudes at the MSTO, see Figure \ref{figure:CMD}). We have also included in the simulated population a binary fraction of 30\% where the mass of the secondary varies uniformly between 0-1 times the mass primary \citep{Keller01}. We then slice the upper MSTO perpendicular to the locus of the MSTO between two bounds as shown in Figure \ref{figure:CMD}. Expressing distance along the axis $X^{\prime}$ we then perform a two-sided KS test on the simulated EMSTO population and a single age (the average of the FG and SG ages) population of identical number. We then determine the visibility of the EMSTO as the fraction of trials in which the EMSTO population exhibits a KS-test probability of distinct parent populations of 95\% or more.

\begin{figure}
\begin{center}
\includegraphics[width=120mm]{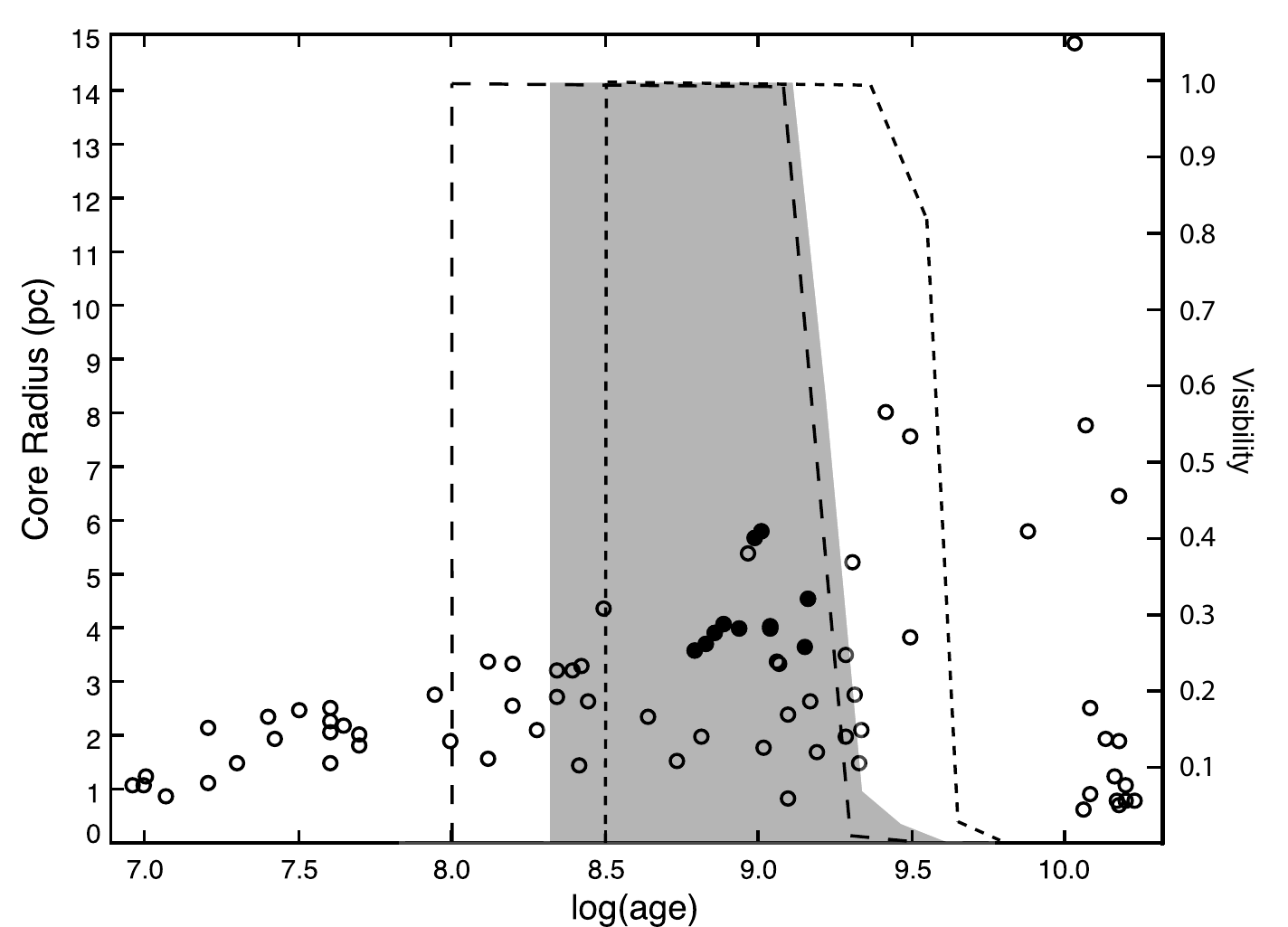}
\caption{The visibility window ({\it{grey shaded region}}) of a simulated EMSTO with an idealised bimodal age distribution with $\Delta$age = 200 Myrs. To younger ages an EMSTO is visible only after 200 Myrs (the formation epoch of the second generation). To older ages the visibility is seen to fall as the age spread becomes a small proportion of the cluster age (see text for details). The dashed and dotted lines are the visibility of a 100 and 300 Myr age spread respectively.}\label{figure:rc_age_monte}
\end{center}
\end{figure}

The results of our EMSTO visibility simulations for a typical LMC cluster as a function of age are shown in Figure \ref{figure:rc_age_monte}. The visibility rapidly falls between log age 9.2 -- 9.4. This drop in visibility is not only related to the decreasing fractional age difference between the FG and SG populations but is also due to a change in the course of stellar evolution at this age. As discussed in \citet{Girardi09} it is at this age that stars transition from possessing enough mass to ignite He in a non-degenerate core, to those that ignite He under degenerate conditions in a He flash at the tip of the RGB. A consequence of this change is that the transition from the MS to RGB is more rapid. Consequently, the extent of the MSTO region is curtailed, and the visibility of the EMSTO is much reduced. In our scenario, clusters such as NGC 1978, 2121, and IC 2146 wouldn't have an observable age spread, even though given their large core radii, they are prime candidates for having progressed recently through the EMSTO phase.

Apart from the extended main-sequence turn-off, there are other features of the CMD that will also reflect the multi-age population. For instance, a red clump extended or bifurcated in luminosity (such as seen in the SMC EMSTO cluster NGC 419 \citeauthor{Girardi09} \citeyear{Girardi09}) or the presence of a broad sub giant branch. However, when in the presence of the background field star population, weakly populated cluster CMD features such as these suffer problematically from field star contamination and represent low contrast effects. The local dominance of the cluster stars in the vicinity of the MSTO remains the key feature for distinguishing the presence of an EMSTO in a particular cluster.

What would such a cluster look like if seen in the process of secondary star formation? Firstly, it is expected that such a cluster will be more massive than those clusters seen in the EMSTO phase now by a factor of 10-20 times \citep[on the basis of gas retention arguments]{Bekki06, dErcole08}. The EMSTO clusters possess typical estimated masses of $log_{10} \rm{M}_{\odot}$ 4.8-5.3 \citet{Mackey03}, hence during secondary star formation they would have amounted to around $10^6$ M$_{\odot}$ systems. As seen in the study of \citet{Mackey03} such clusters represent the extreme high-mass end of the cluster mass function. Secondly, there would be the presence of active star formation that would be revealed by the presence of centrally located young stars, residual gas, and/or a pre-main sequence population (perhaps as emission-line stars). No such progenitor of the EMSTO cluster population is observed: firstly there is only one cluster with a determined mass and age similar to that expected for a EMSTO progenitor (NGC 1856, M$\sim10^{5}$ -- $10^6$ M$_{\odot}$ \citeauthor{Mackey03} \citeyear{Mackey03}, log age=8.12). Furthermore, no known cluster in the age range of log age 8.0-8.5 is known to exhibit a young stellar population, although it most be noted this not a stringent exclusion since present observations would only be able to distinguish a young population of massive stars perhaps in the form of a blue sequence above the MSTO. On the other hand, if there were to be a stepper or truncated IMF for the SG such a population would likely remain hidden to investigations to date.

In considering if the absence of an EMSTO progenitor is a problem for our hypothesis we now consider the evolution of the cluster population in the log age -- N plot of Figure \ref{figure:rc_age}. The number of clusters per interval of log age is approximately constant for ages less than 3 Gyr. If we consider our sample to be largely complete to at least the most massive clusters, this indicates strong evolution occurs in the cluster system, as has been discussed previously (see e.g. \citeauthor{Gieles08} \citeyear{Gieles08}; \citeauthor{Chandar10} \citeyear{Chandar10}).  If there were no preference for the enhanced survivability of EMSTO clusters we would observe $\sim 70$\% of clusters one dex younger as EMSTO cluster progenitors. Instead we see none. This argues that a property (or properties) of EMSTO clusters enhances their ability to survive the physical processes of cluster evolution. The 11 known EMSTO clusters from \citet{Milone09} span an age range of 1.1 Gyr. If no EMSTO cluster was destroyed, and the number of clusters is constant per log age \citep{Chandar10}, we would expect the current population of potential EMSTO progenitor clusters between log age 8.0-8.5 to contain $\sim0.5$ EMSTO progenitors. It is therefore, not problematic to the in situ star formation scenario that we do not see an EMSTO cluster undergoing active star formation in the LMC, there is insufficient cluster formation to sample the upper extent of the cluster mass function in this age range at any given point in time.

It is well understood that only a small fraction of star clusters evolve into long-lived bound systems. In extremely young clusters, the effects of stellar evolution in massive stars (in the form of their supernovae and winds) expel the natal gas of the cluster. With rapid removal of gas, the gravitational potential of the cluster is suddenly reduced and stars that were initially bound to the cluster may now possess velocities in excess of the escape velocity\citep{Decressin10}. This effect accounts for the disruption of up to 90\% of stellar clusters \citep{Lada03}. This phase presumably ends with the removal of the natal gas of the cluster \citep[within 20 Myr, ][]{Goodwin06}. 

The quantitative description of cluster evolution is a topic of debate. \citet{ZhangFall99, Whitmore07, Fall09, Chandar10} determine that in the star cluster system of the merging Antennae galaxies and in the MCs, it is possible to describe the evolution of the cluster system by the bivariate age-mass distribution $g(M,t) = M^{\beta}t^{\gamma}$ where $\beta \approxeq -2$ and $\gamma \approxeq -1$ for clusters up to $\sim1$ Gyr. The result implies that 90\% of clusters are removed for each age dex independent of cluster mass. \citet{Gieles08} challenge that the number count of clusters is likely to be affected by completeness since the sample is essentially luminosity limited rather than mass limited and older clusters are intrinsically fainter. The study of \citeauthor{Gieles08} utilises the mass of the most massive cluster as a function of age to infer the shape of the age distribution. They find $g(M,t) = M^{\beta}t^{\gamma}$ where $\beta \approxeq -2$ and $\gamma \approxeq 0$. This is compatible with {\it{no}} cluster dissolution after the initial gas removal phase until $\sim1$ Gyr, at least for {\it{massive}} clusters. 

The long-term evolution of the cluster luminosity function is likely dominated by the preferential depletion of low-mass clusters by both evaporation due to two-body relaxation and by tidal interactions with the gravitational field of their host galaxy, and the preferential disruption of high-mass clusters by dynamical friction. However, the relative importance of these disruption processes is unclear \citep{Fall01, deGrijs03}. \citet{McLaughlin08} show that they can replicate the Galactic GC mass function, which is characterised by a peak and turnover at high masses, through erosion due to evaporation from two-body internal relaxation of a mass function that initially rose towards lower masses. Increasing cluster mass increases the relaxation time and hence the cluster disruption time. Furthermore, two-body relaxation induced evaporation has a strong and systematic dependance on the GC central density, in the sense that less dense systems suffer stronger depletion. \citet{Elmegreen10} points out the importance of collisions, between dense clouds and other clusters, in the evolution of massive clusters in high density environments at high redshift. This study shows that a power law cluster mass function is rapidly shaped into the log-normal distribution witnessed in the mass function of Milky Way globular clusters. At LMC densities strong collisional dissipation is much reduced, however \citeauthor{Elmegreen10} qualitatively shows that evaporation is sufficient to strongly shape the cluster mass function to leave only the most massive clusters. Furthermore, the morphology of the LMC (namely the lack of a strong bulge) is greatly different to that of the Galaxy and hence the effects of tidal shocks in shaping the stellar cluster luminosity function may be much reduced. In a qualitative sense however, in order to survive to the ancient GC population a star cluster will preferentially be drawn from clusters that are the most massive and the least dense.

\section{Extended main-sequence turn-off clusters as spatially extended clusters}

It is evident in Figures \ref{figure:rc_age} and \ref{figure:rc_rprime} that the EMSTO clusters are, without exception, the most spatially extended clusters for their age and distance from the centre of the LMC. Furthermore, the clusters that exhibit the EMSTO phenomenon most prominently are those which possess the largest core radii. This points to an intriguing connection between the EMSTO phenomenon and the internal {\it{kinematics}} of the clusters in which it is found. In this section we address this finding.

The general form of Figure \ref{figure:rc_age} was first revealed by \citet{Elson87} (with extension in \citeauthor{Elson89} \citeyear{Elson89}, \citeauthor{Elson91} \citeyear{Elson91}, \citeauthor{Elson92} \citeyear{Elson92}, and in the case of the LMC, \citeauthor{Mackey03} \citeyear{Mackey03}). Canonical evolution of an isolated system would lead to an evolutionary path across this figure that (for a Salpeter IMF) rises slightly in core radius over $10^{9.5}$ yr and then declines with the onset of core collapse at $10^{10}$ yr. From a sample of clusters of mass M$>10^{4}$M$_{\odot}$ \citet{Mackey03} shows that no mass--size relation is evident and hence a size-of-sample effect \citep[as discussed in][]{Hunter03} does not explain the radius--age trend that is evident. 

\citet{Mackey08b} address the radius--age trend via $N$-body modelling of massive star clusters analogous to those of the MCs. They identify two processes that can lead to significant and prolonged cluster core expansion, namely, mass-loss due to rapid stellar evolution in a primordially mass-segregated cluster (i.e.\ massive stars are preferentially located in the central regions of the cluster) and heating due to a retained population of stellar mass black holes formed from the SNe of massive stars of the cluster. The timescales over which the two processes dominate differ. The effects of mass-loss in a mass segregated cluster are apparent in the first Gyr. For black hole heating to be important the black holes must dynamically relax to the centre of the cluster a process that takes on order of a Gyr to complete.

In the case of the EMSTO clusters, of ages 600 Myr to 1.5 Gyr, the dominant process responsible for promoting these clusters to larger core radii is therefore that arising from mass loss in an initially mass segregated cluster. \citet{Allison09} and \citet{Moeckel09} show that stellar clusters can dynamically mass segregate on timescales that are much shorter than those expected from consideration of relaxation times. Rather, mass segregation is promoted by the formation of a short-lived, but very dense core that leads to rapid early dynamical evolution. Mass segregation has been reported in a large number of massive clusters in the LMC (e.g. \citet{Gouliermis04}, \citet{Sirianni02}, \citet{deGrijs02}), SMC (e.g.\ NGC330 \citet{Gouliermis04}), and in Galaxy (e.g.\ the Orion nebula cluster, \citeauthor{Preibisch99} \citeyear{Preibisch99}; Westerlund 1, \citeauthor{Gennaro10} \citeyear{Gennaro10}; Arches, \citeauthor{Stolte02} \citeyear{Stolte02}) (see \citeauthor{PortegiesZwart10} \citeyear{PortegiesZwart10} for a review). The detailed simulations of star cluster formation by \citet{Bonnell06}, \citet{McMillan07} and \citet{Moeckel09} are consistent with these observations, suggesting that mass segregation in young massive clusters may well be a product of the formation process, namely, the more massive stars are preferentially formed at the bottom of local potential wells where the gas density is greatest. Furthermore, the degree of mass segregation is expect to be a function of cluster mass - a deeper potential feeds more mass to those central stars \citet{Bonnell06}. In the case of an initially mass segregated cluster \citet{Mackey08b} show that since the massive stars are preferentially concentrated in the central regions of the cluster the effects of early rapid mass-loss due to stellar evolution is greater compared to non-mass segregated counterparts and this results in expansion to much lower central densities. We propose, therefore, that the EMSTO clusters preferentially represent the population of the initially most mass segregated (and hence most massive) clusters to have formed.

As we have discussed above, evolution of the cluster mass function erodes low mass clusters and those clusters of high central density. EMSTO clusters, by virtue of their generally higher mass and lower central densities, preferentially survive to older ages. We propose that clusters that have progressed through the EMSTO phase are the dominant contributors to the ancient GC systems. 

\section{Chemical Abundance Variations}

The scenario we propose predicts that the SG stars within EMSTO clusters will incorporate some proportion of previously processed material from FG stars. This, in turn, leads to the prediction that abundance variations identical in character to those seen in GGCs should be seen in the EMSTO cluster sample. To date, chemical abundance studies of the EMSTO clusters are limited. The chemistry of six stars in the known EMSTO cluster NGC 1783 are presented by \citet{Mucciarelli08}. The six stars do not show an appreciable spread in [O/Fe] or [Na/Fe]. [O/Fe] shows a dispersion of 0.08 dex compared with an expected dispersion from observational uncertainties of 0.14 dex. Similarly, [Na/Fe] shows a dispersion of 0.10 dex compared to 0.11 dex expected. Under our hypothesis, the cluster NGC 1978, due to its large core radius, was highly likely an EMSTO cluster that is now sufficiently old that the EMSTO is no longer visible. However, a similar low abundance variations is seen in this cluster from a sample of 11 stars. 

In the case of the ancient, metal-poor GCs the incorporation of FG processed material is understood to imprint the light element abundance variations clearly evident in the SG population. In the AGB self-pollution scenario, considerable uncertainty remains. Chemical yields based on AGB models by various groups vary dramatically. The models of \citet{Karakas07} obtain Oxygen depletion and Sodium production at levels appropriate to match the O-Na anti-correlation seen in the ancient GC population from the most massive AGBs ($\sim7$M$_{\odot}$). On the other hand, \citet{Ventura08} can replicate the O-Na anti-correlation to considerably lower masses ($\sim5$M$_{\odot}$). Furthermore, differences in the treatment of convection in the models of the above two main modelling groups lead to differences in the level of O depletion. 

The model dependency of Na production and O depletion discussed above similarly clouds our understanding of the yields of AGB stars as a function of metallicity. The models of \citet{Ventura08} extend to [Fe/H]=-0.5 and show that Na production declines with increasing metallicity after a metallicity of [Fe/H] $\sim -1.0$. It is therefore not apparent {\it{a priori}} what quantitative signature we would expect in the case of the EMSTO cluster stars. Furthermore, processed material is incorporated into material of significantly higher metallicities than those of the ancient GC population. A large abundance spread in Na in an ancient GC might be of order +1 dex at [Fe/H] $\sim -1.5$. At a metallicity closer to that of our LMC EMSTO clusters, [Fe/H]$ = -0.5$, the same yield of Na produces only a $+0.3$ dex increase in $A$(Na). Consequently, it is unlikely that the LMC EMSTO clusters would be expected to show the large light element abundance variations seen in the ancient GC population. In the AGB self-pollution scenario, the amount of processed gas is likely to be strongly dependent on the mass of the cluster (see discussion in Section \ref{sect:origins}). The EMSTO clusters of the LMC possess a mean mass of $10^{4.4}$M$_{\odot}$ \citep{Conroy11}, whereas for the GGCs it is $10^{5.1}$M$_{\odot}$ \citep{McLaughlin08}. It is possible therefore that if the level of chemical enrichment is a function of cluster mass that the abundance variations seen in the less massive EMSTO clusters will be lower than those evident in the GGC population.

The expectation of light element abundance variations in the EMSTO population is nonetheless a critical prediction of our scenario and hence warrants further detailed analysis with a significantly enlarged sample of stars from intermediate age clusters in the LMC. It is also clear that further effort on the theoretical yields of AGB stars is required. At metallicities similar to those of the LMC intermediate age clusters are the old GCs NGC 6838 and 47 Tuc ([Fe/H] = -0.83 and -0.77 dex respectively) that possess moderate [O/Na] inter-quartile ranges of 0.257 and 0.472 dex respectively \citep{Carretta10}. The more metal-rich clusters NGC 6388 and NGC 6441 ([Fe/H] = -0.44 and -0.43 respectively) show large inter-quartile ranges in [O/Na]: 0.795 and 0.660 dex respectively \citep{Carretta10}. It is therefore, not clear what expectations should be within the significantly younger LMC clusters. We propose that a study of the chemistry of AGB stars in the young ($\sim 10^8$ yr) clusters of the LMC (for example NGC 1866) could afford useful observational constraints on these models.

\section{Discussion: Origins of the Extended MSTO phenomenon}
\label{sect:origins}

The scenario we present in this paper views the EMSTO clusters as a phase common to those clusters that have been able to survive for a substantial fraction of a Hubble time. We have demonstrated that the apparent restriction of the EMSTO phenomenon to intermediate age clusters is a result of observational selection effects that renders such clusters observationally undetectable at larger ages as the age spread/difference between the multiple/extended populations becomes an insignificant fraction of the cluster age. At earlier times, we have shown that the absence of an EMSTO cluster in the act of formation within the LMC is as expected from a sample size argument - there are insufficient clusters to sample the upper cluster mass range where the EMSTO progenitors reside. Our scenario makes a very clear prediction that the EMSTO clusters should show some light element abundance variations of the same nature as those ubiquitous in the ancient GC population. However, the magnitude of such abundance variations are as yet unclear as they depend upon the yield of a polluting population at considerably higher metallicity than the same polluting population responsible for the ancient GC population. 

Our scenario views the ancient GCs as the distilled remains of an initial star cluster population. The ancient GCs are clusters who, primarily by virtue of their larger mass, have survived to the present day. We hypothesise that by virtue of their larger mass these clusters have been able to retain/acquire gas in order to progress through the EMSTO phase of secondary star formation. That clusters can retain and/or acquire gas for the secondary star formation is an assertion identical to that made in the self-pollution scenario for GCs. It is one that is at first consideration implausible. If the gas required for the formation of the SG was drawn solely from the ejecta of FG stars the amount of gas available would be:

\begin{equation}
M = \int_{M_{t=SG}}^{M_{min SNe}} (m_{initial}-m_{final}) m^{-\alpha} dm
\end{equation}

where $M_{t=SG}$ is the turnoff mass at the time of the formation of the SG, $M_{min SNe}$ is the highest mass that does not expire via supernovae, and $\alpha$ is the initial mass function exponent (Salpeter IMF $\alpha=2.35$). For values of $M_{t=SG} = 4.5$ M$_{\odot}$, $M_{min SNe} = 8$ M$_{\odot}$ and a total cluster mass of $10^{4}$ M$_{\odot}$, we derive a gas mass of $\sim$100 M$_{\odot}$. However, the SG population in every massive cluster studied to date is at least equivalent in number (and hence mass) to the FG, that is on order of 5$\times$10$^3$ M$_{\odot}$ in this example.

It is clear therefore that the self-pollution scenario for EMSTO clusters (and hence by extension the ancient GCs) can not occur without the accumulation of gas from an external source. This is a conclusion drawn by previous studies, for example by \citet{Bekki06} in the case of $\omega$ Cen. By similar mass function arguments as above, this study demonstrated it was not possible to achieve the observationally inferred helium enrichment from the FG. \citet{Bekki06} propose that the SG instead formed from gas ejected from field stellar populations that surrounded $\omega$ Cen when it was the nucleus of a progenitor some 10-20 times more massive than the present-day GC. 

\citet{dErcole08} present simulations which show that gas ejected by FG AGB collects in a cluster core via a cooling flow. The AGB-mass loss plus advected pristine material is then transformed into the SG. Thereafter, \citeauthor{dErcole08} follow the dynamical evolution and mass loss of the composite population. They find that a large fraction of FG stars are lost due to expansion that results from early mass loss produced by FG SNe. The SG population, on the other hand, remains largely unscathed due to its initial concentration in the innermost cluster regions. The resulting cluster after a Hubble time exhibits a ratio of SG to FG stars in line with observations of GGCs. Pristine material is required to fall into the AGB-processed material in order to explain the range in the Na-O anti-correlation seen within individual clusters. The source of the pristine gas is envisaged by \citeauthor{dErcole08} as a shell of material that subsequently falls back onto the dense cloud that is forming SG stars. 

A problem for this scenario is, as \citet{Gratton10} point out, that the pristine gas reservoir associated with the cluster must somehow avoid pollution by SN ejecta. \citeauthor{Gratton10} propose that mass loss from stars below the AGB mass limit could, if they were to loose 0.5-1\% of their mass, provide the pristine gas required. However, as acknowledged by \citeauthor{Gratton10}, this level of mass loss is a somewhat ad-hoc assumption. \citet{Decressin10} examines the dynamical behaviour of a dual population massive cluster in which rapidly rotating stars are the polluters instead of AGBs. As seen in \citet{dErcole08}, if the SG is initially more centrally concentrated, gas expulsion is able to expel FG at the expense of SG stars to produce the inferred fractions in GGCs.

\citet{Conroy11} present arguably the most comprehensive model for the early evolution of massive clusters. This model accounts for the effects of stellar mass loss, SNII and prompt SNIa feedback, ram pressure stripping and accretion from the ambient ISM on the gas content of a GC progenitor. The process outlined by \citeauthor{Conroy11} is as follows: first SNII clear the GC of the initial gas reservoir, mass loss from AGB stars occurs and gas is accreted from the ISM and this material collects at the centre of the cluster potential. At such time the radiation is too intense for the gas to cool. After several $10^8$ yr the radiation intensity declines sufficiently for molecular H to form. This leads to gas cooling and collapse to form a second generation. SNII from the second generation, and prompt SNIa from the FG clear the gas from the cluster to terminate star formation. 

\citeauthor{Conroy11} point out that a critical mass of $\sim 10^4$ M$_{\odot}$ is required to retain gas in the face of ram pressure stripping in the present LMC environment. Above this mass threshold clusters are expected to show multiple populations, and below clusters are predicted to be coeval simple stellar populations. A similar and contemporaneous model for the development of multiple populations with enrichment via rapidly rotating massive stars is presented by \citet{Decressin10}. This study finds that the effect of gas ejection by the FG SNII is to preferentially eject FG stars while SG stars remain bound to the cluster. In the most favourable of cases a SG fraction of 60\% may result. This is a faction in line with observations of both GGCs and the LMC EMSTO clusters. Furthermore, since the SG stars are formed centrally, long term evolution will enhance their fraction.

A series of the most massive Galactic GCs, namely, M54 (apparently the Sagittarius dwarf nuclear GC, \citeauthor{Carretta10c} \citeyear{Carretta10c}), M22 \citep{DaCosta09, Marino09}, and NGC 2419 \citep{Cohen10} show star-to-star iron and/or calcium variations. $\omega$ Cen reveals calcium \citep{Norris96}, iron \citep[amongst other elements,][]{Johnson08} and helium \citep{Norris04, Dupree10} abundance variations together with multiply aged populations \citep[for example,][]{Bedin04}. These clusters are strongly suspected as the nuclear cores of dwarf galaxies accreted to the Milky Way. Under our hypothesis the EMSTO phenomenon occurs over a broad range of cluster mass, from inception at $\sim10^{4}$M$_{\odot}$ in massive clusters to the nuclear star cluster mass regime in excess of $\sim10^{7}$M$_{\odot}$ \citep{Georgiev09}. In the case of a nuclear star cluster, perhaps at the center of its own dark matter halo, the cluster can retain some fraction of the yield from the SNe that it, and surrounding regions, have experienced leading to internal dispersions of Fe-peak and $\alpha$-elements. Constrained by surrounding potential, such nuclear star clusters could undergo multiple cycles of the EMSTO phenomenon of star formation and enrichment. Such evolution would lead to populations of stars with distinct metallicities, chemistry and perhaps ages such as seen in the most massive GCs (see also \citeauthor{Bekki10} \citeyear{Bekki10}). 

\section{Conclusions} 

We have presented a scenario for the formation of extended main-sequence star clusters (EMSTO). In our scenario we envisage the EMSTO phenomenon as arising in an evolutionary phase that is shared by all massive star clusters. The cluster, by virtue of its considerable mass, is able to retain gas expelled from a first generation of stars and accrete gas from the ambient ISM. After 50-300 Myr this gas is sufficiently cool to form a second generation of cluster stars. The chemical signature of the processed material expelled by the first generation of stars is incorporated in the second generation stars. This phase of massive star cluster evolution is therefore the genesis of the light element abundance variations ubiquitous in the ancient globular cluster (GC) population.

The EMSTO clusters represent the missing link in GC evolution between young clusters that have a simple stellar population and the ancient GC that reveal the imprint of multiple stellar populations. The EMSTO phenomenon evident in the LMC is restricted to a narrow age range. We show that this age range over which the phenomenon is observationally apparent is due to selection effects. Amongst $< 10^{8.5}$ yr clusters insufficient clusters are formed in the LMC to stochastically populate the upper portion of the cluster mass function. At older ages, as the age difference between the first and second generation populations becomes an increasingly small fraction of the total cluster age and the phenomenon ceases to be apparent observationally.

We have demonstrated that EMSTO clusters appear as the most extended clusters for their age and distance from the center of the LMC. This indicates that a connection between the kinematics of stars in the cluster and the presence of the EMSTO phenomenon. This connection would not be expected in a scenario in which the EMSTO phenomenon arises from stellar rotation for instance. Rather, we argue that this connection arises naturally from the dynamical evolution of the stellar cluster population. This evolution preferentially preserves massive and initially mass segregated clusters at the expense of low mass clusters to produce the log-normal mass function that characterises the ancient globular cluster population. At intermediate ages those massive and initially mass segregated clusters are the most extended clusters - the EMSTO clusters. In this way, dynamical evolution distils the ancient GC population from the intermediate age EMSTO clusters.

In our proposed scenario, the EMSTO phenomenon establishes the light element abundance variations that are ubiquitous in the ancient GC population. Our scenario therefore makes a very strong prediction that stars in EMSTO clusters should show the light element abundance variations similar in nature to those seen in galactic GCs. However, the expected magnitude of these variations are loosely constrained. The EMSTO clusters of the LMC are at significantly higher metallicities and lower masses than those of the ancient GCs. Considerable uncertainty exists in the yields of stellar evolution at such metallicities (from AGB stars for instance) and the level to which processed material is incorporated in the second generation of stars at lower masses. As a result of the formation scenario we have outlined, we would predict that the second generation of stars in the EMSTO clusters will be more centrally concentrated and show a smaller velocity dispersion than that of the first generation.

\acknowledgments

Thanks to David Yong for valuable discussions over the course of this research. SK acknowledges the financial support of the Australian Research Council through Discovery Project Grant DP0878137. DM acknowledges the financial support from an Australian Research Fellowship (DP1093431) from the Australian Research Council.


%% in parentheses, after the keyword, but they will not be verified.

%{\it Facilities:} \facility{Nickel}, \facility{HST (STIS)}, \facility{CXO (ASIS)}.

%%\bibliographystyle{apj}      

\clearpage

%% the file, uncomment the \plotone, \plottwo, and
%% If you need a layout that cannot be achieved with \plotone or
%% \plottwo, you can invoke the graphicx package directly with the
%% \includegraphics command or use \plotfiddle. For more information,
%% please see the tutorial on "Using Electronic Art with AASTeX" in the
%% documentation section at the AASTeX Web site,
%% http://www.journals.uchicago.edu/AAS/AASTeX.
%%
%% The examples below also include sample markup for submission of
%% supplemental electronic materials. As always, be sure to check
%% the instructions to authors for the journal you are submitting to
%% for specific submissions guidelines as they vary from

%% This example uses \plotone to include an EPS file scaled to
%% 80% of its natural size with \epsscale. Its caption
%% has been written to indicate that additional figure parts will be
%% available in the electronic journal.

\clearpage

\clearpage

\clearpage

\clearpage

%% Here we use \plottwo to present two versions of the same figure,
%% one in black and white for print the other in RGB color
%% for online presentation. Note that the caption indicates
%% that a color version of the figure will be available online.
%%

%% If you are not including electonic art with your submission, you may
%% mark up your captions using the \figcaption command. See the
%% No more than seven \figcaption commands are allowed per page,
%% so if you have more than seven captions, insert a \clearpage

%% Tables should be submitted one per page, so put a \clearpage before

%% Two options are available to the author for producing tables:  the
%% deluxetable environment provided by the AASTeX package or the LaTeX
%% table environment.  Use of deluxetable is preferred.
%%

\end{document}